# Superconductivity in the Hexagonal Ternary Phosphide ScIrP

Y. Okamoto[1,3,*], T. Inohara[1], Y. Yamakawa[2,3], A. Yamakage[1,3], and K. Takenaka[1]

[1]*Department of Applied Physics, Nagoya University, Nagoya 464-8603, Japan*
[2]*Department of Physics, Nagoya University, Nagoya 464-8602, Japan*
[3]*Institute for Advanced Research, Nagoya University, Nagoya 464-8601, Japan*

We report the discovery of a bulk superconducting transition at 3.4 K in the ternary phosphide, ScIrP, which crystallizes in a hexagonal ZrNiAl-type structure without spatial inversion symmetry. On the basis of heat capacity data in a zero magnetic field, ScIrP is suggested to be a weakly-coupled Bardeen-Cooper-Schrieffer superconductor. Alternatively, experimental results under magnetic fields indicate that this material is a type-II superconductor with an upper critical field $H_{c2}$ at magnetic fields above 5 T at zero temperature. This moderately high $H_{c2}$ does not violate the Pauli limit, but it does imply that there is a significant effect from the strong spin-orbit interaction of Ir 5$d$ electrons in the noncentrosymmetric crystal structure. Electronic structure calculations show an interesting feature of ScIrP, where both the Sc 3$d$ and Ir 5$d$ orbitals contribute to the electronic density of states at the Fermi level.

Elemental Ca and Sc show superconductivity with a relatively high transition temperature $T_c$ of ~20 K at high pressures. Ca is not a superconductor at ambient pressure but becomes superconducting above ~50 GPa [1]. $T_c$ increases with increasing pressure and exceeds 20 K at ~100 GPa [2], which is the highest $T_c$ in all elemental metals. Sc exhibits a superconducting transition above ~60 GPa, which approaches 20 K at 107 GPa [3,4]. At higher pressures, however, Sc undergoes a structural phase transition to a nonsuperconducting phase (Sc-III). If this transition did not occur, Sc would exhibit a $T_c$ higher than 20 K. In simple-metal superconductors such as Al, Sn, and Pb, $T_c$ decreases with increasing pressure due to a decrease of the electronic density of states (DOS). The pressure effects on $T_c$ in Ca and Sc, opposite to those in simple metals, have been discussed in terms of the pressure-induced $s$–$d$ electron transfer where 4$s$ electrons are dominant for electron conduction at ambient pressure and are transferred to the narrower 3$d$ band by applying pressure [4,5]. The electrons occupying the 3$d$ band play an important role in relatively high $T_c$ superconductivity.

A similar situation arises at ambient pressure in the graphite intercalation compound CaC$_6$. CaC$_6$ shows a superconducting transition at $T_c$ = 11.5 K, which is the highest $T_c$ in all graphite intercalation compounds and is almost twice that of the second highest ($T_c$ = 6.5 K for YbC$_6$) and two orders of magnitude higher than that of KC$_8$ [6]. This large difference in $T_c$ can be qualitatively understood by the character of conduction electrons. In KC$_8$, YbC$_6$, and CaC$_6$, the C 2$p$, Yb 5$d$, and Ca 3$d$ bands are dominant at the Fermi level, respectively [7]. The higher $T_c$ in CaC$_6$ is most likely caused by a large electronic DOS due to the narrow 3$d$ band, in cooperation with high phonon frequencies derived

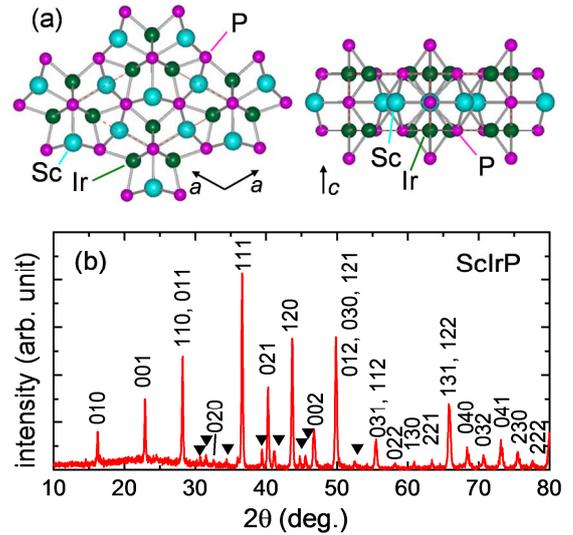

Fig. 1 (a) Crystal structure of ScIrP viewed along (left) and perpendicular (right) to the $c$-axis. The dashed line represents the hexagonal unit cell. (b) A powder XRD pattern of a ScIrP polycrystalline sample taken at room temperature. The peaks indicated by triangles are those of unknown impurities. Peak indices are given using a hexagonal unit cell with lattice constants of $a$ = 6.3309(7) Å and $c$ = 3.8854(4) Å.

from light Ca atoms.

Superconductivity in the 5$d$ band metals of heavy elements such as Ir and Pt usually occurs at lower temperatures. However, the strong spin-orbit interaction of heavy 5$d$ atoms can make it unconventional in contrast to that in the case of light elements. A typical example is the boride superconductor Li$_2$Pt$_3$B ($T_c$ = 3 K) [8]. Li$_2$Pd$_3$B ($T_c$ = 8 K), which is a 4$d$ analogue of Li$_2$Pt$_3$B, is a conventional superconductor with an isotropic gap. However, Li$_2$Pt$_3$B is believed to be a line-nodal superconductor, where spin-triplet states are significantly hybridized with spin-singlet states by an antisymmetric spin-orbit interaction in the noncentrosymmetric crystal structure [9]. Moreover, SrPtAs ($T_c$ = 2.4 K) with a honeycomb structure consisting of Pt and As atoms is theoretically predicted to be a chiral $d$-wave or triplet $f$-wave superconductor, because it crystallizes in a hexagonal and noncentrosymmetric structure [10]. Besides these superconductors, strong-coupling superconductivity has been reported in the antiperovskite phosphide SrPt$_3$P [11] and superconductivity is induced by a structural instability in Ir$_{1-x}$Pt$_x$Te$_2$ [12]. In addition, many interesting Ir- or Pt-based superconductors have been discovered in recent years [13-20]. Thus, 5$d$ electron systems with heavy elements can be viewed as one of the "hot spots" for the search of novel superconductors.

In this Letter, we report on the ternary phosphide ScIrP, a type-II superconductor with $T_c$ = 3.4 K. ScIrP was first synthesized by Phannenschmidt *et al.* in single-crystalline form by the flux method [21]. ScIrP crystallizes in the hexagonal ZrNiAl-type structure with a noncentrosymmetric space group of $P-62m$, as shown in Fig. 1(a). This crystal structure consists of alternately stacked Sc and Ir layers, in which Sc and Ir atoms are square-pyramidally and tetrahedrally coordinated by phosphorous atoms, respectively. In each Sc layer, Sc atoms form a kagome-triangular lattice [22], where Sc$_3$ regular triangles are interconnected in an intermediate geometry between the kagome and triangular lattices. Ir atoms form Ir$_3$ clusters with a regular-triangle shape, which form a triangular lattice in the Ir plane.

There are superconductors with a relatively high $T_c$ such as ZrRuP ($T_c$ = 13.3 K) in the ZrNiAl-type structure [23]. In contrast, Phannenschmidt *et al.* reported that ScIrP does not show a superconducting transition above 2.5 K [21]. We find that ScIrP exhibits a bulk superconducting transition at $T_c$ = 3.4 K using electrical resistivity, magnetization, and heat capacity measurements of polycrystalline samples prepared by a solid-state reaction method. The superconducting properties with zero magnetic field can be understood as a weak-coupling Bardeen-Cooper-Schrieffer (BCS) superconductivity. In contrast, those in a magnetic field are more interesting; the electrical resistivity data obtained under various magnetic fields indicate the presence of a moderately high upper critical field at zero temperature, which may be caused by the strong spin-orbit interaction of Ir 5$d$ electrons in ScIrP.

Polycrystalline samples of ScIrP were prepared by a solid-state reaction method. An equimolar mixture of Sc chips, Ir powder, and black phosphorous powder was sealed in a quartz tube with Ar gas pressure of 0.04-0.05 MPa. The tube was heated to and kept at 673 K for 12 h and at 1173 K for 96 h with an intermediate grinding. The obtained sample was pulverized, pressed into a pellet, wrapped in a Ta foil,

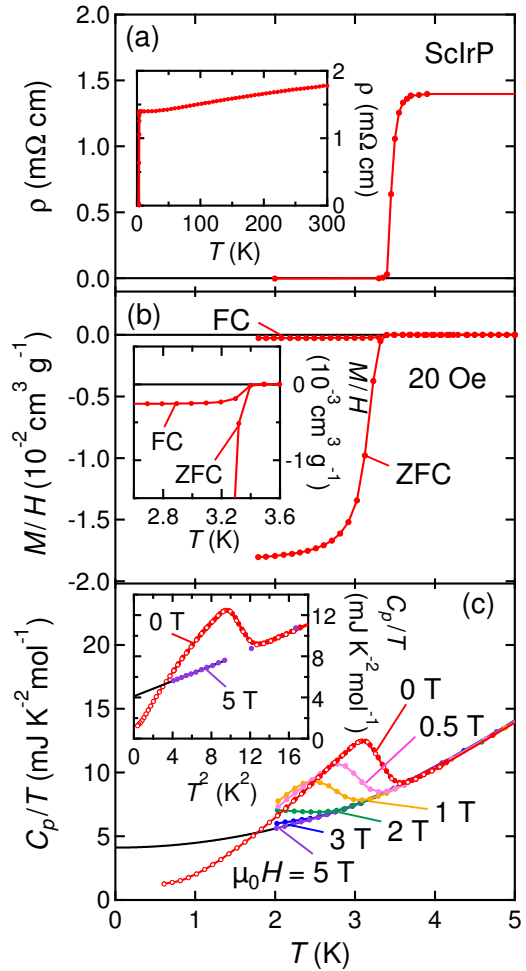

Fig. 2 (a) Temperature dependence of electrical resistivity of a ScIrP polycrystalline sample. The inset shows the data up to room temperature. (b) Temperature dependence of field-cooled and zero-field-cooled magnetizations measured at a magnetic field of 20 Oe. The inset shows an enlarged view of the main panel around the superconducting transition. (c) Temperature dependence of heat capacity divided by temperature, $C_p/T$, as a function of $T$ (main panel) and $T^2$ (inset) measured at various magnetic fields from 0 to 5 T. Filled and open circles represent the experimental data down to 2 and 0.6 K, respectively. The solid line (inset) and curve (main panel) show a fit to the equation $C_p/T = AT^2 + \gamma$ of the 2–3.3 K data taken at $\mu_0 H$ = 5 T.

and then sealed in a quartz tube with Ar gas. The tube was heated and kept at 1423 K for 144 h with intermediate grindings. Sample characterization was performed by powder X-ray diffraction (XRD) analysis with Cu K$\alpha$ radiation at room temperature using a RINT-2100 diffractometer (Rigaku). As shown in Fig. 1(b), all diffraction peaks observed in the powder XRD pattern, with the exception of some small peaks caused by small amounts of unknown impurities, can be indexed on the basis of a hexagonal structure with the lattice constants $a$ = 6.3309(7) Å and $c$ = 3.8854(4) Å. The obtained $a$ and $c$ are ~0.6 and ~0.2% smaller than those reported by Pfannenschmidt *et al.*

[21]. Electrical resistivity, heat capacity, and magnetization measurements were performed using the Physical Property Measurement System and Magnetic Property Measurement System (both from Quantum Design). First principles calculations were performed using the WIEN2k code [24]. Experimental structural parameters were used for the calculations [21].

Figures 2(a) and 2(b) show the temperature dependences of electrical resistivity and magnetization of a ScIrP polycrystalline sample, respectively. Electrical resistivity data show metallic behavior below room temperature and a sharp drop to zero between 3.65 and 3.30 K. A strong diamagnetic signal is observed in the zero-field-cooled magnetization data below 3.4 K, indicating that a bulk superconducting transition occurs at this temperature. Providing that the sample is a single phase of ScIrP, the estimated shielding fraction at 1.8 K is large at 220%, well beyond 100%, which is probably due to the demagnetizing effect. Field-cooled data also show a sharp drop at 3.4 K and a small, but finite Meissner signal below this temperature. These behaviors are typical of a type-II superconductor with pinning. Considering that the midpoint of the resistivity drop, the onset of the magnetization drop, and the onset of the heat capacity jump [Fig. 2(c)] are at 3.46, 3.4, and 3.5 K, respectively, the superconducting transition temperature $T_c$ of ScIrP is determined to be 3.4 K.

The above observations are contradictory to the single-crystal data reported by Pfannenschmidt et al., who claimed that their ScIrP crystals did not show a superconducting transition down to 2.5 K [21]. We think that this contradiction comes from some kind of non-stoichiometry inherent in ScIrP. Generally, the deviation of a chemical composition from the nominal value is usually larger in a single crystal prepared by a flux method than in a polycrystalline sample prepared by a solid state reaction method. As mentioned above, the lattice constants reported by Pfannenschmidt et al. are slightly larger than those obtained in this study. Moreover, recent preliminary experiments indicate that ScIrP samples prepared with an excess of 10 mol% Sc do not exhibit a superconducting transition down to 1.8 K. A small difference in chemical composition may cause the difference in results. As mentioned by Pfannenschmidt et al., however, a quantitative analysis of chemical composition using an energy-dispersive X-ray spectrometer is impossible for ScIrP because the Ir M and P K lines significantly overlap.

Figure 2(c) shows that the heat capacity divided by temperature, $C_p/T$, as a function of $T$ and $T^2$. As shown in the main panel, the zero-field data show a clear jump at 3.0–3.5 K due to the bulk superconducting transition. This jump is shifted to a lower temperature by applying a magnetic field and is suppressed below 2 K at $\mu_0 H = 5$ T. To obtain the lattice heat capacity at low temperatures, the 5 T data between 2.0 and 3.3 K are fitted to the equation $C_p/T =$ $AT^2 + \gamma$, as shown in the inset of Fig. 2(c). $A$ and $\gamma$ represent the coefficient of the $T^3$ term of the lattice heat capacity and the Sommerfeld coefficient in the normal state, respectively. This fit yields $A = 0.372(3)$ mJ K$^{-4}$ mol$^{-1}$ and $\gamma = 4.11(2)$ mJ K$^{-2}$ mol$^{-1}$.

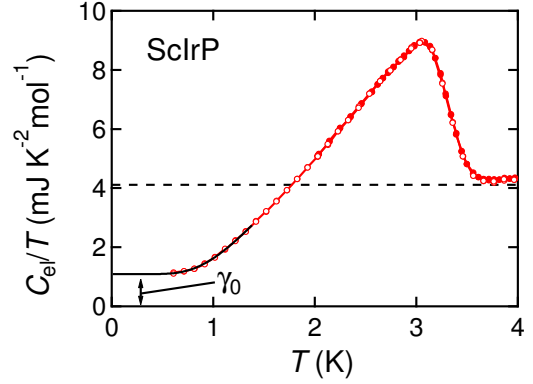

Fig. 3 Temperature dependence of the electron heat capacity divided by temperature, $C_{el}/T$, for ScIrP obtained by subtracting the lattice contribution from the experimental data. The broken line represents the electron heat capacity in the normal state $\gamma$ obtained by the linear fit shown in the inset of Fig. 2(c). The solid curve below $T = 0.4T_c$ shows a fit to the equation $C_{el} = \gamma_0 T + B \exp[-\Delta(0)/k_B T]$.

The electron heat capacity divided by temperature obtained by subtracting the lattice contribution from experimental data, $C_{el}/T = C_p/T - AT^2$, is shown as a function of temperature in Fig. 3. The midpoint of the jump and the width of the superconducting transition are 3.3 and 0.6 K, respectively. $C_{el}/T$ decreases almost linearly below 3 K with decreasing temperature and then approaches a finite $\gamma_0$, i.e., a residual $\gamma$. In Fig. 3, the $C_{el}/T > \gamma$ area is only 10% larger than the $C_{el}/T < \gamma$ one. This means that the entropy balance regarding the superconducting transition is mostly maintained, suggesting that the lattice heat capacity is appropriately estimated. Assuming the presence of an isotropic superconducting gap, the $C_{el}/T$ data below $0.4T_c$ is fitted to the equation $C_{el} = \gamma_0 T + B \exp[-\Delta(0)/k_B T]$, as shown in Fig. 3, where $\Delta(0)$ is a superconducting gap at zero temperature. The result of this fit is reasonably good, yielding $\Delta(0)/k_B = 5.27(9)$ K, $\gamma_0 = 1.09(10)$ mJ K$^{-2}$ mol$^{-1}$, and $B = 103(7)$ mJ K$^{-1}$ mol$^{-1}$. The obtained $\gamma_0$ corresponds to 27% of $\gamma$, which is too large to be attributed to the small amounts of unknown impurities appearing as tiny peaks in Fig. 1(b). It is reasonable to suppose that there is a certain number of nonsuperconducting particles in the ScIrP phase, possibly formed by the inhomogeneity of the chemical composition. The magnitude of the jump at $T_c$ is estimated to be $\Delta C_{el}/(\gamma - \gamma_0)T_c = 1.55$ using $\Delta C_{el}/T_c = 4.7$ mJ K$^{-2}$ mol$^{-1}$ and $\gamma - \gamma_0 = 3.02$ mJ K$^{-2}$ mol$^{-1}$. This value is close to the weak-coupling limit value of 1.43 for a conventional BCS superconductor.

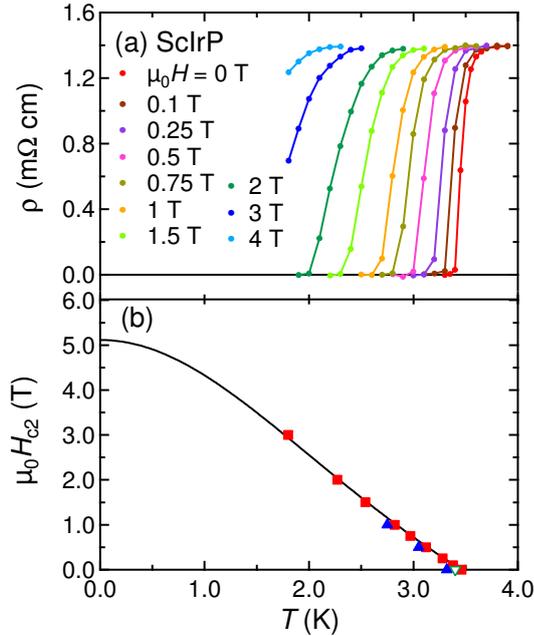

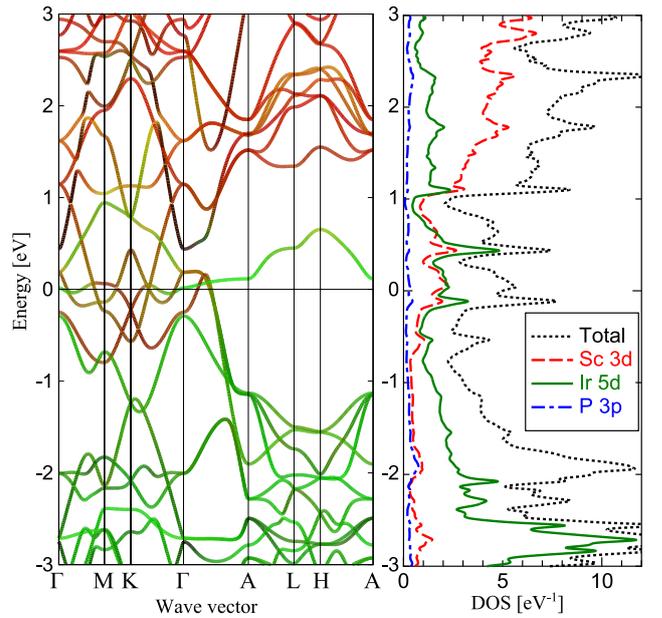

Fig. 4 (a) Temperature dependences of electrical resistivity of a ScIrP polycrystalline sample measured at various magnetic fields of 0–4 T. (b) Temperature dependence of the upper critical field $H_{c2}$. Filled squares, filled triangles, and the open triangle represent $H_{c2}$ determined by the midpoint of the drop in the resistivity data, the midpoint of the jump in the heat capacity data, and the onset of the superconducting transition in the magnetization data, respectively. The solid curve shows a fit to the GL formula of the resistivity data.

Fig. 5 The electronic states of ScIrP without spin-orbit interaction. Electronic band structure (left) and total and partial electronic DOS (right) are shown. The Fermi level is set to 0 eV.

The above results strongly suggest that ScIrP is an *s*-wave superconductor without a node in the superconducting gap. The normalized gap of $2\Delta(0)/k_BT_c = 3.10$, obtained from $\Delta(0)/k_B = 5.27$ K and $T_c = 3.4$ K, is smaller than the theoretical value of $2\Delta(0)/k_BT_c = 3.53$ for weak-coupling BCS superconductivity. Although one may expect the presence of multigap or chiral *d*-wave superconductivity from the small superconducting gap, the lowest measured temperature of 0.6 K (~$0.2T_c$) in this study is not low enough to obtain a conclusive estimation. Future experiments at lower temperatures should uncover the superconducting gap structure of ScIrP.

We now consider the superconducting properties of ScIrP in a magnetic field. Figure 4(a) shows the temperature dependences of electrical resistivity measured at various magnetic fields. The midpoint $T_c$ is suppressed below 1.8 K by applying a relatively large magnetic field of 4 T, suggestive of type-II superconductivity. The upper critical fields, $H_{c2}$, of ScIrP determined using various probes are shown as a function of temperature in Fig. 4(b). The $H_{c2}$ values determined by the electrical resistivity data taken at various magnetic fields are fitted to the Ginzburg-Landau (GL) formula $H_{c2}(T) = H_{c2}(0)[1 - (T/T_c)^2]/[1 + (T/T_c)^2]$, yielding a moderately large $H_{c2}(0)$ of 5.11(4) T and a GL coherence length of $\xi_{GL} = 8.03$ nm. The $H_{c2}(0)$ of ScIrP can be enhanced by the hybridization of spin-triplet states into spin-singlet states due to the significant antisymmetric spin-orbit interaction of the Ir 5*d* orbitals, because there are considerable contributions of Ir 5*d* orbitals in the electronic DOS at the Fermi level, as discussed below. However, the estimated $H_{c2}(0)$ does not exceed the Pauli-limiting field $\mu_0 H_{c2}(0)/\text{T} = 1.84\ T_c/\text{K} = 6.37$ [25], suggestive of the minor contribution of the triplet states in the superconducting phase. For ScIrP, its moderately large $H_{c2}(0)$ may be related to its crystal structure, which has multiple mirror planes as seen in Fig. 1(a), resulting in a weak violation of the inversion symmetry.

Finally, we discuss the electronic structure of the normal state in ScIrP. The electronic band structure and DOS of ScIrP are shown in Fig. 5. The calculated Sommerfeld coefficient $\gamma_{\text{band}} = 3.9$ mJ K$^{-2}$ mol$^{-1}$, estimated from the DOS at the Fermi level (5.0 states eV$^{-1}$, is almost the same as the experimentally obtained $\gamma$ of 4.11 mJ K$^{-2}$ mol$^{-1}$. The weak enhancement of $\gamma$ indicates weak electron-electron and electron-phonon interactions in ScIrP. On the other hand, the magnetic susceptibility $\chi$ at 300 K measured at a magnetic field of 5 T is small ($\chi = 2 \times 10^{-5}$ cm$^3$ mol$^{-1}$). This small $\chi$ probably gives a small Wilson ratio, although the estimation of the Pauli paramagnetic susceptibility is difficult because of the possible presence of a large van Vleck susceptibility and the difficulty in estimating the diamagnetic contribution of core electrons.

An interesting aspect of the electronic structure of ScIrP is that there are significant contributions of both the Sc 3*d* and Ir 5*d* orbitals at the Fermi level. As shown by the partial DOS in the right panel of Fig. 5, the contribution of the Sc $3d$ orbitals is dominant well above the Fermi energy $E_F$, while that of the Ir $5d$ orbitals is well below $E_F$. The orbitals overlap and contribute almost equally to the DOS at around $E_F$. As described at the beginning of this Letter, a metallic system with Sc $3d$ electrons can show a relatively high-$T_c$ superconductivity, while a system with Ir $5d$ electrons is a candidate unconventional superconductor. Even though superconductivity in ScIrP is not observed at a high $T_c$ and can be understood by the conventional BCS mechanism, it should be emphasized that the significant contributions of both the Sc $3d$ and Ir $5d$ orbitals coexist at the Fermi level in this material. This situation is different from other Ir phosphide superconductors [13,18] and may make ScIrP a multigap superconductor, similar to $MgB_2$ [26]. We believe that an unconventional and high-$T_c$ superconductor will be discovered in materials with both $3d$ electrons from a light element and $5d$ electrons from a heavy element.

In summary, ScIrP is found to show a bulk superconducting transition at $T_c$ = 3.4 K, which is the first observation of superconductivity in a Sc phosphide. Superconductivity in ScIrP is suggested to be of the weak-coupling BCS type on the basis of heat capacity measurements. Electronic structure calculations show that there are almost equal contributions of the Sc $3d$ and Ir $5d$ orbitals at the Fermi level. The strong spin-orbit interaction of Ir $5d$ electrons in the noncentrosymmetric crystal structure may give rise to the moderately high $H_{c2}(0)$ in this material.

This work was partly carried out at the Materials Design and Characterization Laboratory under the Visiting Researcher Program of the Institute for Solid State Physics, University of Tokyo and supported by JSPS KAKENHI (Grant Number: 25800188). The authors are grateful to D. Hirai and K. Nawa for their support in heat capacity measurements, to T. Yamauchi for his help in the magnetization measurements, and to Z. Hiroi for helpful discussions.

*e-mail address: yokamoto@nuap.nagoya-u.ac.jp

[1] K. J. Dunn and F. P. Bundy: Phys. Rev. B **25**, 194 (1982); S. Okada, K. Shimizu, T. C. Kobayashi, K. Amaya, and S. Endo: J. Phys. Soc. Jpn. **65**, 1924 (1996).
[2] M. Sakata, Y. Nakamoto, K. Shimizu, T. Matsuoka, and Y. Ohishi: Phys. Rev. B **83**, 220512 (2011).
[3] J. J. Hamlin and J. S. Schilling: Phys. Rev. B **76**, 012505 (2007).
[4] M. Debessai, J. J. Hamlin, and J. S. Schilling: Phys. Rev. B **78**, 064519 (2008).
[5] H. L. Skriver: Phys. Rev. Lett. **49**, 1768 (1982).
[6] T. E. Weller, M. Ellerby, S. S. Saxena, R. P. Smith, and N. T. Skipper: Nat. Phys. **1**, 39 (2005); N. Emery, C. Hérold, M. d'Astuto, V. Garcia, Ch. Bellin, J. F. Marêché, P. Lagrange, and G. Loupias: Phys. Rev. Lett. **95**, 087003 (2005).
[7] G. Csányi, P. B. Littlewood, A. H. Nevidomskyy, C. J. Pickard, and B. D. Simons: Nat. Phys. **1**, 42 (2005); I. I. Mazin: Phys. Rev. Lett. **95**, 227001 (2005); M. Calandra and F. Mauri: Phys. Rev. Lett. **95**, 237002 (2005).
[8] K. Togano, P. Badica, Y. Nakamori, S. Orimo, H. Takeya, and K. Hirata: Phys. Rev. Lett. **93**, 247004 (2004); P. Badica, T. Kondo, and K. Togano: J. Phys. Soc. Jpn. **74**, 1014 (2005).
[9] M. Nishiyama, Y. Inada, and G.-q. Zheng: Phys. Rev. Lett. **98**, 047002 (2007); S. Harada, J. J. Zhou, Y. G. Yao, Y. Inada, and G.-q. Zheng: Phys. Rev. B **86**, 220502 (2012).
[10] Y. Nishikubo, K. Kudo, and M. Nohara: J. Phys. Soc. Jpn. **80**, 055002 (2011); M. H. Fischer, T. Neupert, C. Platt, A. P. Schnyder, W. Hanke, J. Goryo, R. Thomale, and M. Sigrist: Phys. Rev. B **89**, 020509 (2014); J. Goryo, M. H. Fischer, and M. Sigrist: Phys. Rev. B **86**, 100507 (2012).
[11] T. Takayama, K. Kuwano, D. Hirai, Y. Katsura, A. Yamamoto, and H. Takagi: Phys. Rev. Lett. **108**, 237001 (2012).
[12] S. Pyon, K. Kudo, and M. Nohara: J. Phys. Soc. Jpn. **81**, 053701 (2012).
[13] Y. Qi, J. Guo, H. Lei, Z. Xiao, T. Kamiya, and H. Hosono: Phys. Rev. B **89**, 024517 (2014).
[14] G. Eguchi, D. C. Peets, M. Kriener, Y. Maeno, E. Nishibori, Y. Kumazawa, K. Banno, S. Maki, and H. Sawa: Phys. Rev. B **83**, 024512 (2011).
[15] K. Kudo, K. Fujimura, S. Onari, H. Ota, and M. Nohara: Phys. Rev. B **91**, 174514 (2015).
[16] N. Haldolaarachchige, Q. Gibson, L. M. Schoop, H. Luo, and R. J. Cava: J. Phys: Condens. Matter **27**, 185701 (2015).
[17] S. Pyon, K. Kudo, J. Matsumura, H. Ishii, G. Matsuo, M. Nohara, H. Hojo, K. Oka, M. Azuma, V. O. Garlea, K. Kodama, and S. Shamoto: J. Phys. Soc. Jpn. **83**, 093706 (2014); D. Hirai, R. Kawakami, O. V. Magdysyuk, R. E. Dinnebier, A. Yaresko, and H. Takagi: J. Phys. Soc. Jpn. **83**, 103703 (2014).
[18] D. Hirai, T. Takayama, R. Higashinaka, H. A. Katori, and H. Takagi: J. Phys. Soc. Jpn. **78**, 023706 (2009).
[19] T. Shibayama, M. Nohara, H. A. Katori, Y. Okamoto, Z. Hiroi, and H. Takagi: J. Phys. Soc. Jpn. **76**, 073708 (2007).
[20] Y. Qi, S. Matsuishi, J. Guo, H. Mizoguchi, and H. Hosono: Phys. Rev. Lett. **109**, 217002 (2012).
[21] U. Pfannenschmidt, U. Ch. Rodewald, and R. Pöttgen: Z. Naturforsch. **66b**, 205 (2011).
[22] H. Ishikawa, T. Okubo, Y. Okamoto, and Z. Hiroi: J. Phys. Soc. Jpn. **83**, 043703 (2014).
[23] R. Müller, R. N. Shelton, J. W. Richardson, Jr., R. A. Jacobson: J. Less-Common Met. **92**, 177 (1983); G. P Meisner and H. C. Ku: Appl. Phys. A **31**, 201 (1983).
[24] P. Blaha, K. Schwarz, G. Madsen, D. Kvasnicka, and J. Luitz: WIEN2k, An Augmented Plane Wave + Local Orbitals Program for Calculating Crystal Properties (Techn. Universität Wien, Austria, 2001).
[25] A. M. Clogston: Phys. Rev. Lett. **9**, 266 (1962).
[26] P. Szabó, P. Samuely, J. Kačmarčík, T. Klein, J. Marcus, D. Fruchart, S. Miraglia, C. Marcenat, and A. G. M. Jansen: Phys. Rev. Lett. **87**, 137005 (2001).